\documentclass[twocolumn,citeautoscript,prl,superscriptaddress,amsmath,amssymb]{revtex4-1}
 
\usepackage{graphicx}
\usepackage{color}
\usepackage{upgreek}
\usepackage{float}
\usepackage{lipsum}
\usepackage{mathrsfs}
\usepackage{booktabs}
\usepackage{url}
\usepackage{xcolor}
\usepackage[colorlinks,citecolor=blue,linkcolor=blue]{hyperref}
 
\newcommand{\ket}[1]{\ensuremath{\left| #1 \right\rangle}}

\newcommand{\Tr}[1]{\ensuremath{\text{Tr}\left[#1\right]}}

\begin{document}

\title{Filtering crosstalk from bath non-Markovianity via spacetime classical shadows}
% via spacetime classical shadows
\author{G. A. L. White}
\email{greg.white@monash.edu}
\affiliation{School of Physics, University of Melbourne, Parkville, VIC 3010, Australia}
\affiliation{School of Physics and Astronomy, Monash University, Clayton, Victoria 3800, Australia}

\author{K. Modi}
\email{kavan.modi@monash.edu}
\affiliation{School of Physics and Astronomy, Monash University, Clayton, VIC 3800, Australia}
\affiliation{Centre for Quantum Technology, Transport for New South Wales, Sydney, NSW 2000, Australia}

\author{C. D. Hill}
\email{charles.hill1@unsw.edu.au}
\affiliation{School of Physics, University of Melbourne, Parkville, VIC 3010, Australia}
\affiliation{School of Mathematics and Statistics, University of Melbourne, Parkville, VIC, 3010, Australia}
\affiliation{Silicon Quantum Computing, The University of New South Wales, Sydney, New South Wales 2052, Australia}

\begin{abstract}
    From an open system perspective non-Markovian effects due to a nearby bath or neighbouring qubits are dynamically equivalent. However, there is a conceptual distinction to account for: neighbouring qubits may be controlled. 
    We combine recent advances in non-Markovian quantum process tomography with the framework of classical shadows to characterise spatiotemporal quantum correlations. Observables here constitute operations applied to the system, where the free operation is the maximally depolarising channel. Using this as a causal break, we systematically erase causal pathways to narrow down the progenitors of temporal correlations. We show that one application of this is to filter out the effects of crosstalk and probe only non-Markovianity from an inaccessible bath. It also provides a lens on spatiotemporally spreading correlated noise throughout a lattice from common environments. We demonstrate both examples on synthetic data.
    Owing to the scaling of classical shadows, we can erase arbitrarily many neighbouring qubits at no extra cost. Our procedure is thus efficient and amenable to systems even with all-to-all interactions. 
\end{abstract}
\maketitle

\section{Introduction}

\begin{figure}[!t]
  \centering
  \includegraphics[width=0.95\linewidth]{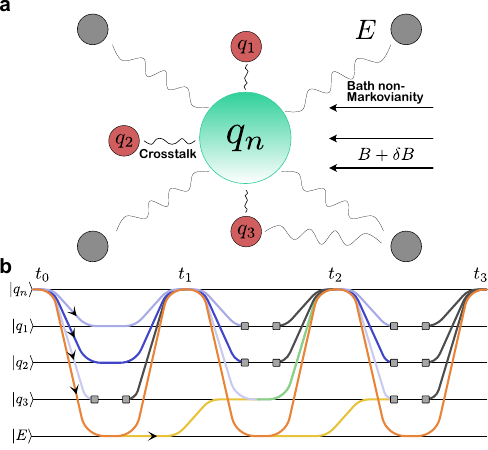}
  \caption{System of interacting qubits and an inaccessible non-Markovian environment $(E)$. \textbf{a} A target qubit $q_n$ may interact via crosstalk mechanisms with other qubits, $\{q_1,q_2,q_3,\cdots\}$, in a quantum device, as well as defects in the bath and fluctuating classical fields $B+\delta B$. \textbf{b} The non-Markovian correlations for that system may be separated into different causal pathways by which the correlations are mediated. Causal breaks (depicted in grey) erase any temporal correlations from a given pathway, allowing one to infer the various contributions to total non-Markovianity from nearby qubits and environment.}
  \label{fig:crosstalk_idea}
\end{figure}

In the race to fault tolerant quantum computing, magnified sensitivity to complex dynamics in open quantum systems requires increasingly tailored characterisation and spectroscopic techniques~\cite{chalermpusitarak2021frame,ferrie2018bayesian,nielsen-gst,PhysRevX.9.021045,White-NM-2020,white2022non,youssry2020characterization,von2020two}. Correlated dynamics are one particularly pernicious class of noise, and can be generated from a variety of sources, including inhomogeneous magnetic fields, coherent bath defects, and nearby qubits, see Figure~\ref{fig:crosstalk_idea}a~\cite{paladino20141,mavadia2018experimental}. Concerningly, these effects are often omitted from %\red{typical}
quantum error correction noise models despite being ubiquitous in noisy intermediate-scale quantum (NISQ) hardware~\cite{Clader2021,White-NM-2020,white2021many, white2022non,correlated-qec,Harper2020}.

Temporal -- or non-Markovian -- correlations are elements of error that are correlated between different points in time, as mediated by interactions with an external system~\cite{Pollock2018a,white2022non}. 
A process is said to be non-Markovian if the total dynamics do not factorise into a product of dynamical maps~\cite{Pollock2018}, a stronger condition than the well-known completely-positive divisibility of dynamics~\cite{milz-divisibility}.
The specific mediator of these effects is both conceptually and experimentally relevant device information. Is it controllable, or is it part of the inaccessible bath? 
Relatedly, if the dynamics of two nearby qubits do not spatially factorise, this is known as crosstalk.
If one qubit is traced out, then entangling crosstalk -- such as the always-on $ZZ$ interactions in transmon qubits~\cite{PhysRevA.101.052308} -- can generate temporal correlations for the second qubit. Whether the dynamics look non-Markovian depends on whether it is feasible or not to dilate the characterisation to multiple qubits and account for the variables responsible for these correlations. Typically, it is not.
Since crosstalk and bath non-Markovianity can easily be conflated, it is crucial to find robust methods that can not only account for their behaviour, but distinguish them.

In this Letter, we establish a systematic, concrete, and efficient approach to the two pragmatic questions: (1) if non-Markovian dynamics are detected across different timescales for a qubit, do they come from neighbouring qubits or a nearby bath? And (2) how can we determine when two qubits are coupled to a shared bath generating common cause non-Markovian effects. %\red{, versus independent sources or direct interaction?}
The solutions here have highly practical implications. Namely, whether curbing the correlated effects is achievable through control or fabrication methods \cite{winick2021simulating,wei2022hamiltonian}.
%\red{Up to this point, attempts to answer the first question have been mostly heuristic: detect non-Markovian behaviour, and then search through candidate explanations of the underlying physics to see whether this explains the phenomena~\cite{tripathi2022suppression,Sarovar2020detectingcrosstalk,krinner2020benchmarking}. But this approach is not scalable, and highly model-dependent.}
Process tensor tomography (PTT) is a recently developed generalisation to quantum process tomography, and can guarantee an answer to these questions and more, but the number of experiments required grows as $\mathcal{O}(d^{2kn+N})$ to find correlations across $k$ steps over $N$ qudits~\cite{white2022non}. \par

The basic premise of our work is to apply the method of classical shadows~\cite{huang-shadow} to PTT, resolving these problems. The classical shadow philosophy implements randomised single-shot measurements to learn properties of a state, granting access to an exponentially larger pool of observables at fixed locality. Employing this, instead of reconstructing the whole multi-time process for an entire quantum register, we can estimate and analyse each of the fixed-weight process marginals. 
Marginalising over a measurement is equivalent to measuring and throwing the outcome away. To marginalise over a process input is equivalent to inputting a maximally mixed state. Hence, these are maximal depolarising channels at no extra cost, which act as causal breaks on controllable systems.

When suitably placed, these operations eliminate temporal correlations as mediated on the chosen Hilbert spaces, thus allowing non-Markovian sources to be causally tested. We illustrate this idea in Figure~\ref{fig:crosstalk_idea}b. The end result is the simultaneous determination of the bath-mediated non-Markovianity on all qubits.
Our approach hence only depends on the individual system size (in this work, qubits), and is a physics-independent way for us to test the relevant hypotheses. 
We are also able to simultaneously compute all spacetime marginals, %\red{i.e. the generic process correlations between groupings of qubit-time coordinates.} 
extending the randomised measurement toolkit to the spatiotemporal domain~\cite{elben2022randomized}. %\red{Note that while this approach can also be used to characterise crosstalk, extensive work has already been conducted on this topic, and so we do not concentrate on these spatial correlations~\cite{Sarovar2020detectingcrosstalk,rudinger2021experimental, helsen2021estimating}. Instead, we focus on the ability to filter out crosstalk effects and study what is left over from the bath.}

\section{Spatiotemporal classical shadows}

By virtue of the state-process equivalence for multi-time processes~\cite{chiribella_memory_2008,Shrapnel_2018,Pollock2018a,PhysRevA.98.012328}, quantum operations on different parts of a system at different times constitute observables on a many-body quantum state. 
This allows state-of-the-art characterisation techniques to be applied to quantum stochastic processes. 
%Consequently, properties of quantum stochastic processes can be similarly probed using state-of-the-art quantum state characterisation techniques. 
Classical shadow tomography~\cite{huang-shadow,elben2022randomized} is one such technique, and already has many generalisations and applications~\cite{helsen2021estimating,hadfield2022measurements,huang2022provably,PhysRevLett.125.200501}. Measuring classical shadows allows for exponentially greater observables to be determined about a state, provided sufficiently low weight. But this restriction
means the technique has limitations for the study of temporal correlations (which are high weight) in contrast to spatial ones, as discussed in Ref.~\cite{white2021many}. %\red{Namely, typical temporal correlations appears in high-weight observables. Hence, the full set of low-weight correlations in-principle detectable by classical shadows is shrunk considerably by conditions of causality on the process. The free operation in classical shadows -- the identity observable -- is a maximally depolarising channel. At a minimum, this limits detections of past-future correlations to those generated by non-unital dynamics, and partially scrambles the environment signal.} 
Our work expands on this to the multi-qubit-multi-time case, and identifies other desirable applications of classical shadows to multi-time processes.

\emph{Definitions and Notation.--}
Consider a quantum device with a register of qudits $\mathbf{Q}:=\{q_1,q_2,\cdots,q_N\}$ across a series of times $\mathbf{T}_k:=\{t_0,t_1,\cdots, t_k\}$. We take the whole quantum device to define the system: $\mathcal{H}_S:=\bigotimes_{j=1}^N\mathcal{H}_{q_j}$. The device interacts with an external, inaccessible environment whose space we denote $\mathcal{H}_E$. % \red{(and interchangeably refer to as a bath)}.
The $k$-step open process is driven by a sequence $\mathbf{A}_{k-1:0}$ of control operations on the whole register, each represented mathematically by completely positive (CP) maps: $\mathbf{A}_{k-1:0} := \{\mathcal{A}_0, \mathcal{A}_1, \cdots, \mathcal{A}_{k-1}\}$, after which one obtains a final state $\rho_k^S(\mathbf{A}_{k-1:0})$ conditioned on this choice of interventions. Note that where we label an object with time information only, that object is assumed to concern the entire register.
These controlled dynamics have the form:
\begin{equation}\label{eq:multiproc}
      \rho_k^S\left(\textbf{A}_{k-1:0}\right) = \text{Tr}_E [U_{k:k-1} \, \mathcal{A}_{k-1} \cdots \, U_{1:0} \, \mathcal{A}_{0} (\rho^{SE}_0)],
\end{equation}
where $U_{k:k-1}(\cdot) = u_{k:k-1} (\cdot) u_{k:k-1}^\dag$. 
Now let the Choi representations of each $\mathcal{A}_j$ be denoted by a caret, i.e. $\hat{\mathcal{A}}_j = \mathcal{A}_j\otimes \mathcal{I}[|\Phi^+\rangle\!\langle \Phi^+|] = \sum_{nm}\mathcal{A}_j[|n\rangle\!\langle m|]\otimes |n\rangle\!\langle m|$.
Then, the driven process in Equation~\eqref{eq:multiproc} for arbitrary $\mathbf{A}_{k-1:0}$ uniquely defines a multi-linear mapping across the register $\mathbf{Q}$ -- called a process tensor, $\Upsilon_{k:0}$ -- via a generalised Born rule~\cite{Pollock2018a,Shrapnel_2018}:

\begin{equation}\label{eq:PToutput}
    p_k^S(\mathbf{A}_{k-1:0})= \text{Tr}  \left[\Upsilon_{k:0}^{}\left(\Pi_k\otimes \hat{\mathcal{A}}_{k-1}\otimes \cdots \hat{\mathcal{A}}_0\right)^\text{T}\right],
\end{equation}
\par 
At each time $t_j$, the process has an output index $\mathfrak{o}_j$ (which is measured), and input index $\mathfrak{i}_{j+1}$ (which feeds back into the process).
The details of process tensors can be found in the appendix, but are not crucial to understanding this work. 
The two important properties that we stress are: (i) a sequence of operations constitutes an observable on the process tensor via Equation~\eqref{eq:PToutput}, generating the connection to classical shadows, and (ii) a process tensor forms a collection of possibly correlated completely positive, trace-preserving (CPTP) maps, and hence may be marginalised in both time and space to yield the $j$th CPTP map describing the dynamics of the $i$th qubit $\hat{\mathcal{E}}^{(q_i)}_{j:j-1}$.
A process is said to be Markovian if and only if its process tensor is a product state across time.
The measure of non-Markovianity we use throughout this work is that described in Ref.~\cite{Pollock2018}. Specifically, it is the relative entropy $S[\rho\|\sigma] = \Tr{\rho (\log\rho - \log\sigma)}$ between a process tensor $\Upsilon_{k:0}^{}$ and its closest Markov description, the product of its marginals:
\begin{equation}
    \Upsilon_{k:0}^{\text{(Markov)}} = \hat{\mathcal{E}}_{k:k-1}^{}\otimes \cdots \otimes \hat{\mathcal{E}}_{1:0}^{}\otimes \rho_0^{}.
\end{equation}
We denote this generalised quantum mutual information (QMI) for a given process by $\mathcal{N}(\Upsilon_{k:0})$. 
Classical shadow tomography provides access to a small number of low weight observables, with $\langle\mathbb{I}\rangle$ on the remainder of the subsystems. The case where $\langle\mathbb{I}_{\mathfrak{i}_{j+1}}\mathbb{I}_{\mathfrak{o}_j}\rangle$ is evaluated is equivalent to selecting an $\hat{\mathcal{A}}_j =\mathbb{I}_{\mathfrak{i}_{j+1}}\otimes\mathbb{I}_{\mathfrak{o}_j}\equiv \mathbb{I}_{\mathfrak{i}_{j+1}\mathfrak{o}_j}$. This is the Choi state of the maximal depolarising channel, up to normalisation.
When marginalising across all but a handful of times or qubits, we will denote the remaining steps or registers by commas, i.e.
\begin{equation}
\label{PT-marg}
\begin{split}
    \hat{\mathcal{E}}^{(q_{i_0},q_{i_1})}_{j_0:j_0-1,j_1:j_1-1}\!&:=\! \text{Tr}_{ {\{\overline{q_{i_0}},\overline{q_{i_1}}\}}, \{\overline{t_{j_0}},\overline{t_{j_0-1}},\overline{t_{j_1}},\overline{t_{j_1-1}}\}}[\Upsilon_{k:0}];\\
    \Upsilon_{k:0}^{(q_i)} &= \text{Tr}_{\overline{q_i}}[\Upsilon_{k:0}],
    \end{split}
\end{equation}
where the overlines denote complement, i.e. every qubit except $q_i$ : $\mathbf{Q}\backslash \{q_i\}$, or every time except $t_j: \mathbf{T}_k\backslash\{t_j\}$.\par

When non-Markovian correlations persist as mediated by the inaccessible bath, we designate this as bath non-Markovianity (BNM). When the correlations are mediated from neighbouring qubits, we designate this as register non-Markovianity (RNM).
Naturally, since the bath cannot be controlled by definition, BNM can be probed without RNM, but RNM effects cannot be isolated by themselves. Instead, one might consider the spatial process marginals alone to measure direct crosstalk~\cite{Sarovar2020detectingcrosstalk,rudinger2021experimental,helsen2021estimating}.
\par

\begin{figure}
    \centering
    \includegraphics[width=\linewidth]{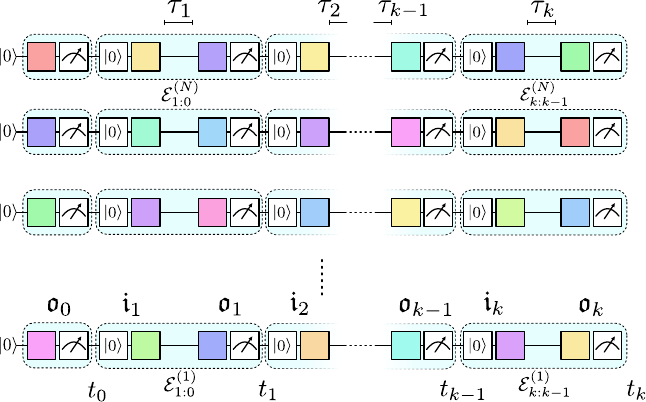}
    \caption{Circuit diagram of our proposed procedure. Spatiotemporal classical shadows can be obtained by applying random Clifford operations to each qubit, projectively measuring, resetting, and then a random Clifford preparation. By repeating these instruments across the circuit with chosen wait times, the appropriate shadow post-processing may be used to determine spatiotemporal marginals of the process. Each $\mathfrak{o}_j$ signifies an index where the state of the system is read out at time $t_j$, and each $\mathfrak{i}_{j+1}$ signifies the preparation of a new state, also at time $t_j$.}
    \label{fig:shadow-circ}
\end{figure}

\emph{Procedure.--}
To map these correlations on each qubit, the classical shadows procedure naturally extends as follows: at each $t\in \mathbf{T}_k$, on each $q\in \mathbf{Q}$, apply a unitary operation randomly selected from the single qubit Clifford group, followed by a projective measurement in the $Z$-basis. This defines a POVM on all qubits across all times $\{U_i^\dagger |x\rangle\!\langle x| U_i\}_{t_j}^{q_m}$.
The measurements considered have four defining features: the qubit $q$ on which they act, the time $t$ at which they are implemented, and the basis change $U$ applied prior to a measurement outcome $x$. To avoid notational overload, we omit these final two labels when writing instruments where the context is clear.

Record both the outcome of the measurement and the random unitary. Reset the qubit to state $\ket{0}$ and apply a random Clifford gate, recording this operation as well. The intended effect of this is to apply a randomised quantum instrument -- i.e. a random measurement with an independently random post-measurement state. 
See Figure~\ref{fig:shadow-circ} for the circuit diagram. %\red{Note that any informationally complete set of instruments here is valid, but we choose measure-and-prepare operations for conceptual ease. }\par 
The application of an instrument in each chosen location in space and time constitutes a single-shot piece of information about the process tensor. The single shot is a projection of the process tensor onto the sequence of interleaving measurements $\Pi$ and preparations $P$:
\begin{equation}
    \begin{split}\hat{\mathbf{\Pi}}_{\mathbf{T}_k} &=  \bigotimes_{l=0}^k \bigotimes_{j=1}^N (U_{\mathfrak{o}_{l}}^{j\ast} |x\rangle\!\langle x | U_{\mathfrak{o}_{l}}^{j\text{T}}),\\
        \hat{\mathbf{P}}_{\mathbf{T}_{k}} &=\bigotimes_{l=1}^k \bigotimes_{j=1}^N(U_{\mathfrak{i}_l}^j |0\rangle\!\langle 0 | U_{\mathfrak{i}_l}^{j\dagger}),
    \end{split}
\end{equation}
with probability given in accordance with Equation~\eqref{eq:PToutput}. Note that the Choi state of a POVM element is given by its transpose~\cite{Milz2021PRXQ}: $\hat{\Pi} = \Pi^{\text{T}}$.
Using the tensor product structure, we can examine each measurement and each preparation at each time on each qubit separately. The preparations $P_l^{q_j}$ are all deterministic, and enact the quantum channel
\begin{equation}
    \mathcal{M}_{\mathcal{P}}(\sigma_{\mathfrak{i}_l}^{q_j}) = \mathbb{E}_{U_{\mathfrak{i}_l}^j\sim \mathcal{U}}[U_{\mathfrak{i}_l}^j|0\rangle\!\langle 0|U_{\mathfrak{i}_l}^{j\dagger}],
\end{equation}
where $\mathbb{E}_{U_{\mathfrak{i}_l}^j\sim \mathcal{U}}$ is the expectation value taken over the unitary ensemble.
The inverse of this gives the classical shadow on the process input legs
\begin{equation}
    \hat{D}_{\mathfrak{i}_l}^{j} := \mathcal{M}_\mathcal{P}^{-1}(U^j_{\mathfrak{i}_l}|0\rangle\!\langle 0|U^{j\dagger}_{\mathfrak{i}_l}) = 3 U^j_{\mathfrak{i}_l}|0\rangle\!\langle 0|U^{j\dagger}_{\mathfrak{i}_l} - \mathbb{I}.
\end{equation}
Existence of this inverse is guaranteed by tomographic completeness of the ensemble~\cite{huang-shadow}.
For the measurements $\Pi_l^{q_j}$ we have the usual single qubit Clifford channel:
\begin{equation}
    \mathcal{M}_{\mathcal{D}}(\sigma_{\mathfrak{o}_l}^{q_j}) = \mathbb{E}_{U_{\mathfrak{o}_l}^j\sim \mathcal{U}, x\sim \text{Tr}[\Pi_l^{q_j}\sigma_{\mathfrak{o}_l}^{q_j}]}[U_{\mathfrak{o}_l}^{j\ast}|x\rangle\!\langle x|U_{\mathfrak{o}_l}^{j\text{T}}].
\end{equation}
Here, $|x\rangle$ on each qubit at each time is sampled according to the generalised Born rule in Equation~\eqref{eq:PToutput}, and depends generally on the operations that come before it. The inverse of this channel gives the shadow on the output legs:
\begin{equation}
\hat{\Delta}_{\mathfrak{o}_l}^j:= \mathcal{M}_{\mathcal{D}}^{-1}(U_{\mathfrak{o}_l}^{j\ast}|x\rangle\!\langle x|U_{\mathfrak{o}_l}^{j\text{T}}) = 3U_{\mathfrak{o}_l}^{j\ast}|x\rangle\!\langle x|U_{\mathfrak{o}_l}^{j\text{T}} - \mathbb{I}.
\end{equation}

Hence, for a $k$-step process on $N$ qubits, the classical shadow is a reshuffling of
\begin{equation}
\label{eq:process-shadow}
    \begin{split}
        \hat{\Upsilon}_{k:0} &= \hat{\mathbf{D}}_{\mathbf{T}_k}^{\text{T}}\otimes \hat{\mathbf{\Delta}}_{\mathbf{T}_k}^{\text{T}},
    \end{split}
\end{equation}
to have the $\mathfrak{o}$ and $\mathfrak{i}$ legs alternating, and from which properties can be efficiently determined using the usual median-of-means estimation described in Ref.~\cite{huang-shadow}.

\begin{figure}
    \centering
    \includegraphics[width=\linewidth]{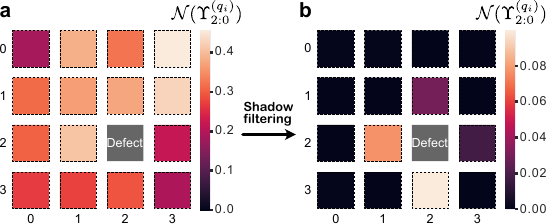}
    \caption{A numerical simulation demonstrating the proposed technique to isolate environmental effects. A grid of 15 qubits is simulated with crosstalk effects and an inaccessible non-Markovian defect. \textbf{a} Determining the non-Markovianity on each qubit individually (with the remainder idle) is not very informative, since each qubit looks non-Markovian due to passive crosstalk. \textbf{b} After learning all of the shadow marginals $\Upsilon_{2:0}^{(q_j)}$, the crosstalk is filtered out to reveal which qubits possess temporal correlations from the environment.}
    \label{fig:simulated-baths}
\end{figure}
\section{Erasing non-Markovian pathways}
The above procedure suffices to estimate marginals of a process tensor with only logarithmic overhead, which we show for completeness in the appendix. In short, we estimate the required observables to uniquely fix the process marginal, and then employ a maximum likelihood algorithm to determine a physically consistent process tensor. %\red{This comes equipped with the familiar logarithmic scaling properties of classical shadows in the number of marginals.}
We consider two possible applications of spatiotemporal classical shadows, supplemented by numerical demonstrations.

\emph{Distinguishing between passive crosstalk and bath non-Markovianity.--}
First, we consider certifying when non-Markovian correlations originate via an inaccessible bath, or from neighbouring qubits in the register.
Certifying bath non-Markovianity means estimating $\Upsilon_{k:0}^{(q_i)}$ -- the marginal process tensor for a single qubit. %\red{Through the usual procedure of classical shadows,} %t
This can be simultaneously performed for all $q_i\in \mathbf{Q}$. The Choi state of the operations on the remainder of the qubits at each time will be $\mathbb{I}/d$, i.e. a maximally depolarising channel. Because this enacts a causal break %\red{-- the operation is a product state -- }
any information travelling from the system into the register cannot persist forwards in time. Hence, computing $\mathcal{N}(\Upsilon_{k:0}^{(q_i)})$ will be a measure of correlations from an inaccessible bath alone. We formally show this in the appendix. 

\begin{figure}[b]
    \centering
    \includegraphics[width=\linewidth]{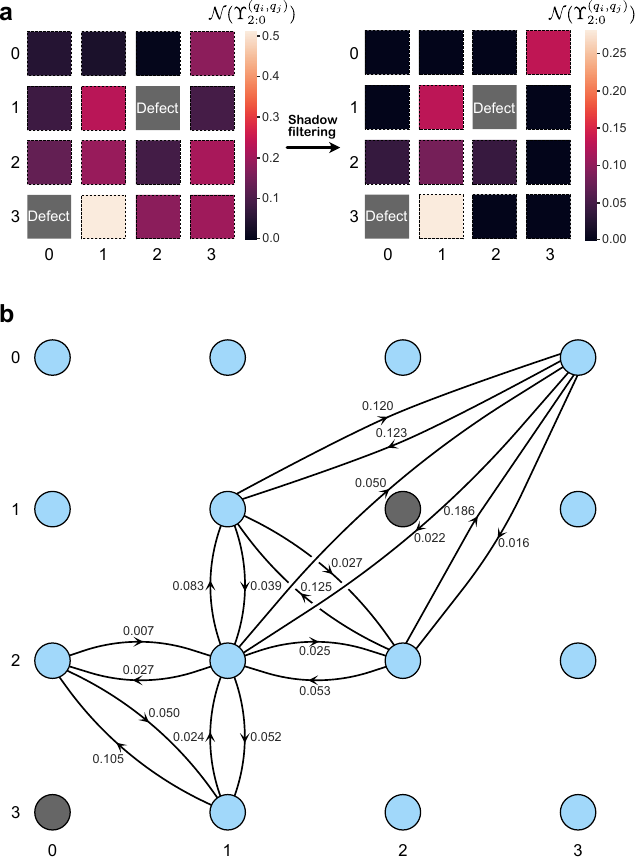}
    \caption{
    A numerical simulation demonstrating the proposed technique to determine qubits with shared baths. A grid of 14 qubits is simulated with crosstalk effects and two inaccessible non-Markovian defects. \textbf{a} Shadow filtering may be used to find qubits coupled to an inaccessible bath as before
    \textbf{b} By looking at the correlations between map $\hat{\mathcal{E}}_{1:0}^{(q_m)}$ with $\hat{\mathcal{E}}_{2:1}^{(q_n)}$, we can infer which qubits share common baths and the extent to which the defects redistribute quantum information.
    Note, the relationship lines between qubits are not direct crosstalk interactions, but bath-mediated spatiotemporal correlation.}
    \label{fig:cc-relations}
\end{figure}

We demonstrate this numerically in Figure~\ref{fig:simulated-baths}. Here, we have 15 qubits and one defect quantum system acting as the bath in a two-step process, and then compute $\mathcal{N}(\Upsilon_{2:0}^{q_i})$. The qubits each experience a random nearest-neighbour $ZZ$-coupling crosstalk, and the ones geometrically closest to the defect are Heisenberg-coupled to that system: $H = \sum_{i}\sum_{\alpha}J_{i,E}^{\alpha}\sigma_{\alpha}^{(i)}\sigma_{\alpha}^{(E)}$ for random $J_{i,E}^{\alpha}$. Figure~\ref{fig:simulated-baths}a shows the standard fare: estimating the process tensor of each qubit and determining its non-Markovianity while the other qubits remain idle. However, the results are not so informative, because they do not distinguish between RNM and BNM effects, and so every qubit experiences temporal correlations. 
Figure~\ref{fig:simulated-baths}b shows the results of a shadow marginal estimation, and
we readily identify only the qubits coupled to bath defects have a non-zero $\mathcal{N}(\Upsilon_{2:0}^{(q_i)})$.

\emph{Identifying shared baths.--}
A second important scenario we consider is where two qubits are correlated via common cause from a shared bath. For example, this might be experiencing the same stray magnetic field inhomogeneities or through a coupling to a common defect.
This is sometimes referred to as crosstalk, because the joint map $\hat{\mathcal{E}}_{j:j-1}^{(q_1,q_2)}$ does not factor to $\hat{\mathcal{E}}_{j:j-1}^{(q_1)}\otimes \hat{\mathcal{E}}_{j:j-1}^{(q_2)}$~\cite{Sarovar2020detectingcrosstalk}. However, we consider this a coarse description because neither system acts as a direct cause for each other's dynamics. Instead, they are subject to spatiotemporal correlations as mediated by the same non-Markovian bath. %\red{ -- i.e. the bath is the common cause for the shared correlations.} 
The key, therefore, is to measure the relationship between the maps $\hat{\mathcal{E}}_{j:j-1}^{(q_1)}$ and $\hat{\mathcal{E}}_{j+1:j}^{(q_2)}$.

We demonstrate this numerically in Figure~\ref{fig:cc-relations}. We have a similar setup to before, except this time with two bath defects. Performing a shadow filtering (Figure~\ref{fig:cc-relations}a) again reveals which qubits are coupled to the defects. However, in Figure~\ref{fig:cc-relations}b, we look at the spacetime marginals estimated from the shadow data. This fine-grained data indicates which qubits are commonly coupled to bath defects, versus independently coupled. The arrows from qubit $q_i$ to qubit $q_j$ indicate the QMI in the process with marginals $\hat{\mathcal{E}}_{2:1}^{(q_j)}\otimes \Upsilon_{1:0}^{(q_i)}$, where $\Upsilon_{1:0}$ also includes initial correlations.
The erasure of $q_j$ in the first step and $q_i$ in the second step eliminates the possibility of direct-cause correlations between the two qubits, leaving only the possibility of common-cause. In other words, non-zero values are a measure of non-Markovian correlations distributed by a shared bath between two qubits. This generates a more informative view of the connected interplay between qubits and their environment%\red{, breaking down the extent to which different systems share temporal correlations and which are fully independent from one another. }

\section{Discussion}
We have introduced a scalable and conceptually simple method to distinguish between non-Markovian dynamics generated by nearby qubits in a quantum device, and those from an inaccessible bath. 
This contributes to the growing zoo of quantum benchmarking techniques, and yet satisfies a unique niche. Geometrically isolating non-Markovian sources across a device can inform various facets of the development process: the signals can warrant further investigation and inform the fabrication process; flag qubits to be given extra control attention; and be fed forward to error-correction decoders.

This also extends the capabilities of the randomised measurement toolbox to the multi-time and multi-qubit domain, and we %Despite the limitations of classical shadows for quantum stochastic processes, we 
have identified an important use case in efficient casual testing.
We anticipate that there exist many alternate applications of classical shadows to spatiotemporal quantum states beyond what we have discussed here. 
Notably, classical shadows have seen extensive recent generalisation and application to quantum speed-up in the determination of quantum properties~\cite{doi:10.1126/science.abn7293,PhysRevLett.126.190505,huang2022provably,elben2022randomized}. Dynamic sampling of small systems, meanwhile, has been shown to be complex in the multi-time sampling setting~\cite{aloisio-complexity}. 
We provide a template by which a similar approach may be applied to quantum stochastic processes. The learning of spatiotemporal correlations constitutes the most general platform for this task, combining many-body states with multi-time processes.\par
Further research is needed to explore alternative ensembles suitable for multi-time processes. Specifically, whether it is tractable to efficiently learn global observables; which properties provide useful information about the non-Markovian interactions; and whether causality conditions imply a learnability gap between quantum states and quantum processes.

We note also that we have introduced our filter in full generality with respect to PTT, but the techniques are generic: the only important point is that causal breaks are applied to neighbouring qubits between each steps. 
The same principles will broadly apply to other approaches to learning non-Markovian dynamics~\cite{PhysRevLett.124.140502,PhysRevLett.112.110401,nielsen-gst,PRXQuantum.2.040351}.

\begin{acknowledgments}
\emph{Acknowledgments.--} 
This work was supported by the University of Melbourne through the establishment of an IBM Quantum Network Hub at the University.
G.A.L.W. is supported by an Australian Government Research Training Program Scholarship. 
C.D.H. is supported through a Laby Foundation grant at The University of Melbourne. 
K.M. is supported through Australian Research Council Discovery Project DP22010179.
K.M. and C.D.H. acknowledge the support of Australian Research Council's Discovery Project DP210100597.
K.M. and C.D.H. were recipients of the International Quantum U Tech Accelerator award by the US Air Force Research Laboratory.

\end{acknowledgments}

\section*{References}
%\bibliography{references}

\begin{thebibliography}{39}%
\makeatletter
\providecommand \@ifxundefined [1]{%
 \@ifx{#1\undefined}
}%
\providecommand \@ifnum [1]{%
 \ifnum #1\expandafter \@firstoftwo
 \else \expandafter \@secondoftwo
 \fi
}%
\providecommand \@ifx [1]{%
 \ifx #1\expandafter \@firstoftwo
 \else \expandafter \@secondoftwo
 \fi
}%
\providecommand \natexlab [1]{#1}%
\providecommand \enquote  [1]{``#1''}%
\providecommand \bibnamefont  [1]{#1}%
\providecommand \bibfnamefont [1]{#1}%
\providecommand \citenamefont [1]{#1}%
\providecommand \href@noop [0]{\@secondoftwo}%
\providecommand \href [0]{\begingroup \@sanitize@url \@href}%
\providecommand \@href[1]{\@@startlink{#1}\@@href}%
\providecommand \@@href[1]{\endgroup#1\@@endlink}%
\providecommand \@sanitize@url [0]{\catcode `\\12\catcode `\$12\catcode
  `\&12\catcode `\#12\catcode `\^12\catcode `\_12\catcode `\%12\relax}%
\providecommand \@@startlink[1]{}%
\providecommand \@@endlink[0]{}%
\providecommand \url  [0]{\begingroup\@sanitize@url \@url }%
\providecommand \@url [1]{\endgroup\@href {#1}{\urlprefix }}%
\providecommand \urlprefix  [0]{URL }%
\providecommand \Eprint [0]{\href }%
\providecommand \doibase [0]{http://dx.doi.org/}%
\providecommand \selectlanguage [0]{\@gobble}%
\providecommand \bibinfo  [0]{\@secondoftwo}%
\providecommand \bibfield  [0]{\@secondoftwo}%
\providecommand \translation [1]{[#1]}%
\providecommand \BibitemOpen [0]{}%
\providecommand \bibitemStop [0]{}%
\providecommand \bibitemNoStop [0]{.\EOS\space}%
\providecommand \EOS [0]{\spacefactor3000\relax}%
\providecommand \BibitemShut  [1]{\csname bibitem#1\endcsname}%
\let\auto@bib@innerbib\@empty
%</preamble>
\bibitem [{\citenamefont {Chalermpusitarak}\ \emph {et~al.}(2021)\citenamefont
  {Chalermpusitarak}, \citenamefont {Tonekaboni}, \citenamefont {Wang},
  \citenamefont {Norris}, \citenamefont {Viola},\ and\ \citenamefont
  {Paz-Silva}}]{chalermpusitarak2021frame}%
  \BibitemOpen
  \bibfield  {author} {\bibinfo {author} {\bibfnamefont {T.}~\bibnamefont
  {Chalermpusitarak}}, \bibinfo {author} {\bibfnamefont {B.}~\bibnamefont
  {Tonekaboni}}, \bibinfo {author} {\bibfnamefont {Y.}~\bibnamefont {Wang}},
  \bibinfo {author} {\bibfnamefont {L.~M.}\ \bibnamefont {Norris}}, \bibinfo
  {author} {\bibfnamefont {L.}~\bibnamefont {Viola}}, \ and\ \bibinfo {author}
  {\bibfnamefont {G.~A.}\ \bibnamefont {Paz-Silva}},\ }\href {\doibase
  10.1103/PRXQuantum.2.030315} {\bibfield  {journal} {\bibinfo  {journal} {PRX
  Quantum}\ }\textbf {\bibinfo {volume} {2}},\ \bibinfo {pages} {030315}
  (\bibinfo {year} {2021})}\BibitemShut {NoStop}%
\bibitem [{\citenamefont {Ferrie}\ \emph {et~al.}(2018)\citenamefont {Ferrie},
  \citenamefont {Granade}, \citenamefont {Paz-Silva},\ and\ \citenamefont
  {Wiseman}}]{ferrie2018bayesian}%
  \BibitemOpen
  \bibfield  {author} {\bibinfo {author} {\bibfnamefont {C.}~\bibnamefont
  {Ferrie}}, \bibinfo {author} {\bibfnamefont {C.}~\bibnamefont {Granade}},
  \bibinfo {author} {\bibfnamefont {G.}~\bibnamefont {Paz-Silva}}, \ and\
  \bibinfo {author} {\bibfnamefont {H.~M.}\ \bibnamefont {Wiseman}},\ }\href
  {https://doi.org/10.1088/1367-2630/aaf207} {\bibfield  {journal} {\bibinfo
  {journal} {New Journal of Physics}\ }\textbf {\bibinfo {volume} {20}},\
  \bibinfo {pages} {123005} (\bibinfo {year} {2018})}\BibitemShut {NoStop}%
\bibitem [{\citenamefont {Nielsen}\ \emph {et~al.}(2020)\citenamefont
  {Nielsen}, \citenamefont {Gamble}, \citenamefont {Rudinger}, \citenamefont
  {Scholten}, \citenamefont {Young},\ and\ \citenamefont
  {Blume-Kohout}}]{nielsen-gst}%
  \BibitemOpen
  \bibfield  {author} {\bibinfo {author} {\bibfnamefont {E.}~\bibnamefont
  {Nielsen}}, \bibinfo {author} {\bibfnamefont {J.~K.}\ \bibnamefont {Gamble}},
  \bibinfo {author} {\bibfnamefont {K.}~\bibnamefont {Rudinger}}, \bibinfo
  {author} {\bibfnamefont {T.}~\bibnamefont {Scholten}}, \bibinfo {author}
  {\bibfnamefont {K.}~\bibnamefont {Young}}, \ and\ \bibinfo {author}
  {\bibfnamefont {R.}~\bibnamefont {Blume-Kohout}},\ }\href
  {http://arxiv.org/abs/2009.07301} {\bibfield  {journal} {\bibinfo  {journal}
  {arXiv:2009.07301}\ } (\bibinfo {year} {2020})}\BibitemShut {NoStop}%
\bibitem [{\citenamefont {Rudinger}\ \emph {et~al.}(2019)\citenamefont
  {Rudinger}, \citenamefont {Proctor}, \citenamefont {Langharst}, \citenamefont
  {Sarovar}, \citenamefont {Young},\ and\ \citenamefont
  {Blume-Kohout}}]{PhysRevX.9.021045}%
  \BibitemOpen
  \bibfield  {author} {\bibinfo {author} {\bibfnamefont {K.}~\bibnamefont
  {Rudinger}}, \bibinfo {author} {\bibfnamefont {T.}~\bibnamefont {Proctor}},
  \bibinfo {author} {\bibfnamefont {D.}~\bibnamefont {Langharst}}, \bibinfo
  {author} {\bibfnamefont {M.}~\bibnamefont {Sarovar}}, \bibinfo {author}
  {\bibfnamefont {K.}~\bibnamefont {Young}}, \ and\ \bibinfo {author}
  {\bibfnamefont {R.}~\bibnamefont {Blume-Kohout}},\ }\href {\doibase
  10.1103/PhysRevX.9.021045} {\bibfield  {journal} {\bibinfo  {journal} {Phys.
  Rev. X}\ }\textbf {\bibinfo {volume} {9}},\ \bibinfo {pages} {021045}
  (\bibinfo {year} {2019})}\BibitemShut {NoStop}%
\bibitem [{\citenamefont {White}\ \emph {et~al.}(2020)\citenamefont {White},
  \citenamefont {Hill}, \citenamefont {Pollock}, \citenamefont {Hollenberg},\
  and\ \citenamefont {Modi}}]{White-NM-2020}%
  \BibitemOpen
  \bibfield  {author} {\bibinfo {author} {\bibfnamefont {G.~A.~L.}\
  \bibnamefont {White}}, \bibinfo {author} {\bibfnamefont {C.~D.}\ \bibnamefont
  {Hill}}, \bibinfo {author} {\bibfnamefont {F.~A.}\ \bibnamefont {Pollock}},
  \bibinfo {author} {\bibfnamefont {L.~C.~L.}\ \bibnamefont {Hollenberg}}, \
  and\ \bibinfo {author} {\bibfnamefont {K.}~\bibnamefont {Modi}},\ }\href
  {\doibase 10.1038/s41467-020-20113-3} {\bibfield  {journal} {\bibinfo
  {journal} {Nature Communications}\ }\textbf {\bibinfo {volume} {11}},\
  \bibinfo {pages} {6301} (\bibinfo {year} {2020})},\ \Eprint
  {http://arxiv.org/abs/2004.14018} {arXiv:2004.14018} \BibitemShut {NoStop}%
\bibitem [{\citenamefont {White}\ \emph {et~al.}(2022)\citenamefont {White},
  \citenamefont {Pollock}, \citenamefont {Hollenberg}, \citenamefont {Modi},\
  and\ \citenamefont {Hill}}]{white2022non}%
  \BibitemOpen
  \bibfield  {author} {\bibinfo {author} {\bibfnamefont {G.~A.~L.}\
  \bibnamefont {White}}, \bibinfo {author} {\bibfnamefont {F.~A.}\ \bibnamefont
  {Pollock}}, \bibinfo {author} {\bibfnamefont {L.~C.~L.}\ \bibnamefont
  {Hollenberg}}, \bibinfo {author} {\bibfnamefont {K.}~\bibnamefont {Modi}}, \
  and\ \bibinfo {author} {\bibfnamefont {C.~D.}\ \bibnamefont {Hill}},\ }\href
  {\doibase https://doi.org/10.1103/PRXQuantum.3.020344} {\bibfield  {journal}
  {\bibinfo  {journal} {PRX Quantum}\ }\textbf {\bibinfo {volume} {3}},\
  \bibinfo {pages} {020344} (\bibinfo {year} {2022})}\BibitemShut {NoStop}%
\bibitem [{\citenamefont {Youssry}\ \emph {et~al.}(2020)\citenamefont
  {Youssry}, \citenamefont {Paz-Silva},\ and\ \citenamefont
  {Ferrie}}]{youssry2020characterization}%
  \BibitemOpen
  \bibfield  {author} {\bibinfo {author} {\bibfnamefont {A.}~\bibnamefont
  {Youssry}}, \bibinfo {author} {\bibfnamefont {G.~A.}\ \bibnamefont
  {Paz-Silva}}, \ and\ \bibinfo {author} {\bibfnamefont {C.}~\bibnamefont
  {Ferrie}},\ }\href {https://doi.org/10.1038/s41534-020-00332-8} {\bibfield
  {journal} {\bibinfo  {journal} {npj Quantum Information}\ }\textbf {\bibinfo
  {volume} {6}},\ \bibinfo {pages} {95} (\bibinfo {year} {2020})}\BibitemShut
  {NoStop}%
\bibitem [{\citenamefont {von L\"upke}\ \emph {et~al.}(2020)\citenamefont {von
  L\"upke}, \citenamefont {Beaudoin}, \citenamefont {Norris}, \citenamefont
  {Sung}, \citenamefont {Winik}, \citenamefont {Qiu}, \citenamefont
  {Kjaergaard}, \citenamefont {Kim}, \citenamefont {Yoder}, \citenamefont
  {Gustavsson}, \citenamefont {Viola},\ and\ \citenamefont
  {Oliver}}]{von2020two}%
  \BibitemOpen
  \bibfield  {author} {\bibinfo {author} {\bibfnamefont {U.}~\bibnamefont {von
  L\"upke}}, \bibinfo {author} {\bibfnamefont {F.}~\bibnamefont {Beaudoin}},
  \bibinfo {author} {\bibfnamefont {L.~M.}\ \bibnamefont {Norris}}, \bibinfo
  {author} {\bibfnamefont {Y.}~\bibnamefont {Sung}}, \bibinfo {author}
  {\bibfnamefont {R.}~\bibnamefont {Winik}}, \bibinfo {author} {\bibfnamefont
  {J.~Y.}\ \bibnamefont {Qiu}}, \bibinfo {author} {\bibfnamefont
  {M.}~\bibnamefont {Kjaergaard}}, \bibinfo {author} {\bibfnamefont
  {D.}~\bibnamefont {Kim}}, \bibinfo {author} {\bibfnamefont {J.}~\bibnamefont
  {Yoder}}, \bibinfo {author} {\bibfnamefont {S.}~\bibnamefont {Gustavsson}},
  \bibinfo {author} {\bibfnamefont {L.}~\bibnamefont {Viola}}, \ and\ \bibinfo
  {author} {\bibfnamefont {W.~D.}\ \bibnamefont {Oliver}},\ }\href {\doibase
  10.1103/PRXQuantum.1.010305} {\bibfield  {journal} {\bibinfo  {journal} {PRX
  Quantum}\ }\textbf {\bibinfo {volume} {1}},\ \bibinfo {pages} {010305}
  (\bibinfo {year} {2020})}\BibitemShut {NoStop}%
\bibitem [{\citenamefont {Paladino}\ \emph {et~al.}(2014)\citenamefont
  {Paladino}, \citenamefont {Galperin}, \citenamefont {Falci},\ and\
  \citenamefont {Altshuler}}]{paladino20141}%
  \BibitemOpen
  \bibfield  {author} {\bibinfo {author} {\bibfnamefont {E.}~\bibnamefont
  {Paladino}}, \bibinfo {author} {\bibfnamefont {Y.~M.}\ \bibnamefont
  {Galperin}}, \bibinfo {author} {\bibfnamefont {G.}~\bibnamefont {Falci}}, \
  and\ \bibinfo {author} {\bibfnamefont {B.~L.}\ \bibnamefont {Altshuler}},\
  }\href {\doibase 10.1103/RevModPhys.86.361} {\bibfield  {journal} {\bibinfo
  {journal} {Rev. Mod. Phys.}\ }\textbf {\bibinfo {volume} {86}},\ \bibinfo
  {pages} {361} (\bibinfo {year} {2014})}\BibitemShut {NoStop}%
\bibitem [{\citenamefont {Mavadia}\ \emph {et~al.}(2018)\citenamefont
  {Mavadia}, \citenamefont {Edmunds}, \citenamefont {Hempel}, \citenamefont
  {Ball}, \citenamefont {Roy}, \citenamefont {Stace},\ and\ \citenamefont
  {Biercuk}}]{mavadia2018experimental}%
  \BibitemOpen
  \bibfield  {author} {\bibinfo {author} {\bibfnamefont {S.}~\bibnamefont
  {Mavadia}}, \bibinfo {author} {\bibfnamefont {C.}~\bibnamefont {Edmunds}},
  \bibinfo {author} {\bibfnamefont {C.}~\bibnamefont {Hempel}}, \bibinfo
  {author} {\bibfnamefont {H.}~\bibnamefont {Ball}}, \bibinfo {author}
  {\bibfnamefont {F.}~\bibnamefont {Roy}}, \bibinfo {author} {\bibfnamefont
  {T.}~\bibnamefont {Stace}}, \ and\ \bibinfo {author} {\bibfnamefont
  {M.}~\bibnamefont {Biercuk}},\ }\href
  {https://doi.org/10.1038/s41534-017-0052-0} {\bibfield  {journal} {\bibinfo
  {journal} {npj Quantum information}\ }\textbf {\bibinfo {volume} {4}},\
  \bibinfo {pages} {7} (\bibinfo {year} {2018})}\BibitemShut {NoStop}%
\bibitem [{\citenamefont {Clader}\ \emph {et~al.}(2021)\citenamefont {Clader},
  \citenamefont {Trout}, \citenamefont {Barnes}, \citenamefont {Schultz},
  \citenamefont {Quiroz},\ and\ \citenamefont {Titum}}]{Clader2021}%
  \BibitemOpen
  \bibfield  {author} {\bibinfo {author} {\bibfnamefont {B.~D.}\ \bibnamefont
  {Clader}}, \bibinfo {author} {\bibfnamefont {C.~J.}\ \bibnamefont {Trout}},
  \bibinfo {author} {\bibfnamefont {J.~P.}\ \bibnamefont {Barnes}}, \bibinfo
  {author} {\bibfnamefont {K.}~\bibnamefont {Schultz}}, \bibinfo {author}
  {\bibfnamefont {G.}~\bibnamefont {Quiroz}}, \ and\ \bibinfo {author}
  {\bibfnamefont {P.}~\bibnamefont {Titum}},\ }\href {\doibase
  10.1103/PhysRevA.103.052428} {\bibfield  {journal} {\bibinfo  {journal}
  {Physical Review A}\ }\textbf {\bibinfo {volume} {103}},\ \bibinfo {pages}
  {052428} (\bibinfo {year} {2021})},\ \Eprint
  {http://arxiv.org/abs/2101.11631} {arXiv:2101.11631} \BibitemShut {NoStop}%
\bibitem [{\citenamefont {White}\ \emph {et~al.}(2021)\citenamefont {White},
  \citenamefont {Pollock}, \citenamefont {Hollenberg}, \citenamefont {Hill},\
  and\ \citenamefont {Modi}}]{white2021many}%
  \BibitemOpen
  \bibfield  {author} {\bibinfo {author} {\bibfnamefont {G.~A.~L.}\
  \bibnamefont {White}}, \bibinfo {author} {\bibfnamefont {F.~A.}\ \bibnamefont
  {Pollock}}, \bibinfo {author} {\bibfnamefont {L.~C.~L.}\ \bibnamefont
  {Hollenberg}}, \bibinfo {author} {\bibfnamefont {C.~D.}\ \bibnamefont
  {Hill}}, \ and\ \bibinfo {author} {\bibfnamefont {K.}~\bibnamefont {Modi}},\
  }\href {http://arxiv.org/abs/2107.13934} {\bibfield  {journal} {\bibinfo
  {journal} {arXiv:2107.13934}\ } (\bibinfo {year} {2021})}\BibitemShut
  {NoStop}%
\bibitem [{\citenamefont {Nickerson}\ and\ \citenamefont
  {Brown}(2019)}]{correlated-qec}%
  \BibitemOpen
  \bibfield  {author} {\bibinfo {author} {\bibfnamefont {N.~H.}\ \bibnamefont
  {Nickerson}}\ and\ \bibinfo {author} {\bibfnamefont {B.~J.}\ \bibnamefont
  {Brown}},\ }\href {\doibase 10.22331/q-2019-04-08-131} {\bibfield  {journal}
  {\bibinfo  {journal} {Quantum}\ }\textbf {\bibinfo {volume} {3}},\ \bibinfo
  {pages} {131} (\bibinfo {year} {2019})},\ \Eprint
  {http://arxiv.org/abs/1712.00502} {arXiv:1712.00502} \BibitemShut {NoStop}%
\bibitem [{\citenamefont {Harper}\ \emph {et~al.}(2020)\citenamefont {Harper},
  \citenamefont {Flammia},\ and\ \citenamefont {Wallman}}]{Harper2020}%
  \BibitemOpen
  \bibfield  {author} {\bibinfo {author} {\bibfnamefont {R.}~\bibnamefont
  {Harper}}, \bibinfo {author} {\bibfnamefont {S.~T.}\ \bibnamefont {Flammia}},
  \ and\ \bibinfo {author} {\bibfnamefont {J.~J.}\ \bibnamefont {Wallman}},\
  }\href {\doibase 10.1038/s41567-020-0992-8} {\bibfield  {journal} {\bibinfo
  {journal} {Nature Physics}\ }\textbf {\bibinfo {volume} {16}},\ \bibinfo
  {pages} {1184} (\bibinfo {year} {2020})},\ \Eprint
  {http://arxiv.org/abs/1907.13022} {arXiv:1907.13022} \BibitemShut {NoStop}%
\bibitem [{\citenamefont {Pollock}\ \emph
  {et~al.}(2018{\natexlab{a}})\citenamefont {Pollock}, \citenamefont
  {Rodr{\'{i}}guez-Rosario}, \citenamefont {Frauenheim}, \citenamefont
  {Paternostro},\ and\ \citenamefont {Modi}}]{Pollock2018a}%
  \BibitemOpen
  \bibfield  {author} {\bibinfo {author} {\bibfnamefont {F.~A.}\ \bibnamefont
  {Pollock}}, \bibinfo {author} {\bibfnamefont {C.}~\bibnamefont
  {Rodr{\'{i}}guez-Rosario}}, \bibinfo {author} {\bibfnamefont
  {T.}~\bibnamefont {Frauenheim}}, \bibinfo {author} {\bibfnamefont
  {M.}~\bibnamefont {Paternostro}}, \ and\ \bibinfo {author} {\bibfnamefont
  {K.}~\bibnamefont {Modi}},\ }\href {\doibase 10.1103/PhysRevA.97.012127}
  {\bibfield  {journal} {\bibinfo  {journal} {Physical Review A}\ }\textbf
  {\bibinfo {volume} {97}},\ \bibinfo {pages} {012127} (\bibinfo {year}
  {2018}{\natexlab{a}})},\ \Eprint {http://arxiv.org/abs/1512.00589}
  {arXiv:1512.00589} \BibitemShut {NoStop}%
\bibitem [{\citenamefont {Pollock}\ \emph
  {et~al.}(2018{\natexlab{b}})\citenamefont {Pollock}, \citenamefont
  {Rodr{\'{i}}guez-Rosario}, \citenamefont {Frauenheim}, \citenamefont
  {Paternostro},\ and\ \citenamefont {Modi}}]{Pollock2018}%
  \BibitemOpen
  \bibfield  {author} {\bibinfo {author} {\bibfnamefont {F.~A.}\ \bibnamefont
  {Pollock}}, \bibinfo {author} {\bibfnamefont {C.}~\bibnamefont
  {Rodr{\'{i}}guez-Rosario}}, \bibinfo {author} {\bibfnamefont
  {T.}~\bibnamefont {Frauenheim}}, \bibinfo {author} {\bibfnamefont
  {M.}~\bibnamefont {Paternostro}}, \ and\ \bibinfo {author} {\bibfnamefont
  {K.}~\bibnamefont {Modi}},\ }\href {\doibase 10.1103/PhysRevLett.120.040405}
  {\bibfield  {journal} {\bibinfo  {journal} {Physical Review Letters}\
  }\textbf {\bibinfo {volume} {120}},\ \bibinfo {pages} {040405} (\bibinfo
  {year} {2018}{\natexlab{b}})},\ \Eprint {http://arxiv.org/abs/1801.09811}
  {arXiv:1801.09811} \BibitemShut {NoStop}%
\bibitem [{\citenamefont {Milz}\ \emph {et~al.}(2019)\citenamefont {Milz},
  \citenamefont {Kim}, \citenamefont {Pollock},\ and\ \citenamefont
  {Modi}}]{milz-divisibility}%
  \BibitemOpen
  \bibfield  {author} {\bibinfo {author} {\bibfnamefont {S.}~\bibnamefont
  {Milz}}, \bibinfo {author} {\bibfnamefont {M.~S.}\ \bibnamefont {Kim}},
  \bibinfo {author} {\bibfnamefont {F.~A.}\ \bibnamefont {Pollock}}, \ and\
  \bibinfo {author} {\bibfnamefont {K.}~\bibnamefont {Modi}},\ }\href {\doibase
  10.1103/PhysRevLett.123.040401} {\bibfield  {journal} {\bibinfo  {journal}
  {Physical Review Letters}\ }\textbf {\bibinfo {volume} {123}},\ \bibinfo
  {pages} {040401} (\bibinfo {year} {2019})},\ \Eprint
  {http://arxiv.org/abs/1901.05223} {arXiv:1901.05223} \BibitemShut {NoStop}%
\bibitem [{\citenamefont {Magesan}\ and\ \citenamefont
  {Gambetta}(2020)}]{PhysRevA.101.052308}%
  \BibitemOpen
  \bibfield  {author} {\bibinfo {author} {\bibfnamefont {E.}~\bibnamefont
  {Magesan}}\ and\ \bibinfo {author} {\bibfnamefont {J.~M.}\ \bibnamefont
  {Gambetta}},\ }\href {\doibase 10.1103/PhysRevA.101.052308} {\bibfield
  {journal} {\bibinfo  {journal} {Phys. Rev. A}\ }\textbf {\bibinfo {volume}
  {101}},\ \bibinfo {pages} {052308} (\bibinfo {year} {2020})}\BibitemShut
  {NoStop}%
\bibitem [{\citenamefont {Winick}\ \emph {et~al.}(2021)\citenamefont {Winick},
  \citenamefont {Wallman},\ and\ \citenamefont
  {Emerson}}]{winick2021simulating}%
  \BibitemOpen
  \bibfield  {author} {\bibinfo {author} {\bibfnamefont {A.}~\bibnamefont
  {Winick}}, \bibinfo {author} {\bibfnamefont {J.~J.}\ \bibnamefont {Wallman}},
  \ and\ \bibinfo {author} {\bibfnamefont {J.}~\bibnamefont {Emerson}},\ }\href
  {\doibase 10.1103/PhysRevLett.126.230502} {\bibfield  {journal} {\bibinfo
  {journal} {Phys. Rev. Lett.}\ }\textbf {\bibinfo {volume} {126}},\ \bibinfo
  {pages} {230502} (\bibinfo {year} {2021})}\BibitemShut {NoStop}%
\bibitem [{\citenamefont {Wei}\ \emph {et~al.}(2022)\citenamefont {Wei},
  \citenamefont {Magesan}, \citenamefont {Lauer}, \citenamefont {Srinivasan},
  \citenamefont {Bogorin}, \citenamefont {Carnevale}, \citenamefont {Keefe},
  \citenamefont {Kim}, \citenamefont {Klaus}, \citenamefont {Landers},
  \citenamefont {Sundaresan}, \citenamefont {Wang}, \citenamefont {Zhang},
  \citenamefont {Steffen}, \citenamefont {Dial}, \citenamefont {McKay},\ and\
  \citenamefont {Kandala}}]{wei2022hamiltonian}%
  \BibitemOpen
  \bibfield  {author} {\bibinfo {author} {\bibfnamefont {K.~X.}\ \bibnamefont
  {Wei}}, \bibinfo {author} {\bibfnamefont {E.}~\bibnamefont {Magesan}},
  \bibinfo {author} {\bibfnamefont {I.}~\bibnamefont {Lauer}}, \bibinfo
  {author} {\bibfnamefont {S.}~\bibnamefont {Srinivasan}}, \bibinfo {author}
  {\bibfnamefont {D.~F.}\ \bibnamefont {Bogorin}}, \bibinfo {author}
  {\bibfnamefont {S.}~\bibnamefont {Carnevale}}, \bibinfo {author}
  {\bibfnamefont {G.~A.}\ \bibnamefont {Keefe}}, \bibinfo {author}
  {\bibfnamefont {Y.}~\bibnamefont {Kim}}, \bibinfo {author} {\bibfnamefont
  {D.}~\bibnamefont {Klaus}}, \bibinfo {author} {\bibfnamefont
  {W.}~\bibnamefont {Landers}}, \bibinfo {author} {\bibfnamefont
  {N.}~\bibnamefont {Sundaresan}}, \bibinfo {author} {\bibfnamefont
  {C.}~\bibnamefont {Wang}}, \bibinfo {author} {\bibfnamefont {E.~J.}\
  \bibnamefont {Zhang}}, \bibinfo {author} {\bibfnamefont {M.}~\bibnamefont
  {Steffen}}, \bibinfo {author} {\bibfnamefont {O.~E.}\ \bibnamefont {Dial}},
  \bibinfo {author} {\bibfnamefont {D.~C.}\ \bibnamefont {McKay}}, \ and\
  \bibinfo {author} {\bibfnamefont {A.}~\bibnamefont {Kandala}},\ }\href
  {\doibase 10.1103/PhysRevLett.129.060501} {\bibfield  {journal} {\bibinfo
  {journal} {Phys. Rev. Lett.}\ }\textbf {\bibinfo {volume} {129}},\ \bibinfo
  {pages} {060501} (\bibinfo {year} {2022})}\BibitemShut {NoStop}%
\bibitem [{\citenamefont {Huang}\ \emph {et~al.}(2020)\citenamefont {Huang},
  \citenamefont {Kueng},\ and\ \citenamefont {Preskill}}]{huang-shadow}%
  \BibitemOpen
  \bibfield  {author} {\bibinfo {author} {\bibfnamefont {H.-Y.}\ \bibnamefont
  {Huang}}, \bibinfo {author} {\bibfnamefont {R.}~\bibnamefont {Kueng}}, \ and\
  \bibinfo {author} {\bibfnamefont {J.}~\bibnamefont {Preskill}},\ }\href
  {\doibase https://doi.org/10.1038/s41567-020-0932-7} {\bibfield  {journal}
  {\bibinfo  {journal} {Nature Physics}\ }\textbf {\bibinfo {volume} {16}},\
  \bibinfo {pages} {1050} (\bibinfo {year} {2020})},\ \Eprint
  {http://arxiv.org/abs/2002.08953} {arXiv:2002.08953} \BibitemShut {NoStop}%
\bibitem [{\citenamefont {Elben}\ \emph {et~al.}(2022)\citenamefont {Elben},
  \citenamefont {Flammia}, \citenamefont {Huang}, \citenamefont {Kueng},
  \citenamefont {Preskill}, \citenamefont {Vermersch},\ and\ \citenamefont
  {Zoller}}]{elben2022randomized}%
  \BibitemOpen
  \bibfield  {author} {\bibinfo {author} {\bibfnamefont {A.}~\bibnamefont
  {Elben}}, \bibinfo {author} {\bibfnamefont {S.~T.}\ \bibnamefont {Flammia}},
  \bibinfo {author} {\bibfnamefont {H.-Y.}\ \bibnamefont {Huang}}, \bibinfo
  {author} {\bibfnamefont {R.}~\bibnamefont {Kueng}}, \bibinfo {author}
  {\bibfnamefont {J.}~\bibnamefont {Preskill}}, \bibinfo {author}
  {\bibfnamefont {B.}~\bibnamefont {Vermersch}}, \ and\ \bibinfo {author}
  {\bibfnamefont {P.}~\bibnamefont {Zoller}},\ }\href
  {https://arxiv.org/abs/2203.11374} {\bibfield  {journal} {\bibinfo  {journal}
  {arXiv preprint arXiv:2203.11374}\ } (\bibinfo {year} {2022})}\BibitemShut
  {NoStop}%
\bibitem [{\citenamefont {Chiribella}\ \emph {et~al.}(2008)\citenamefont
  {Chiribella}, \citenamefont {D'Ariano},\ and\ \citenamefont
  {Perinotti}}]{chiribella_memory_2008}%
  \BibitemOpen
  \bibfield  {author} {\bibinfo {author} {\bibfnamefont {G.}~\bibnamefont
  {Chiribella}}, \bibinfo {author} {\bibfnamefont {G.~M.}\ \bibnamefont
  {D'Ariano}}, \ and\ \bibinfo {author} {\bibfnamefont {P.}~\bibnamefont
  {Perinotti}},\ }\href {\doibase 10.1103/PhysRevLett.101.180501} {\bibfield
  {journal} {\bibinfo  {journal} {Physical Review Letters}\ }\textbf {\bibinfo
  {volume} {101}},\ \bibinfo {pages} {180501} (\bibinfo {year}
  {2008})}\BibitemShut {NoStop}%
\bibitem [{\citenamefont {Shrapnel}\ \emph {et~al.}(2018)\citenamefont
  {Shrapnel}, \citenamefont {Costa},\ and\ \citenamefont
  {Milburn}}]{Shrapnel_2018}%
  \BibitemOpen
  \bibfield  {author} {\bibinfo {author} {\bibfnamefont {S.}~\bibnamefont
  {Shrapnel}}, \bibinfo {author} {\bibfnamefont {F.}~\bibnamefont {Costa}}, \
  and\ \bibinfo {author} {\bibfnamefont {G.}~\bibnamefont {Milburn}},\ }\href
  {\doibase 10.1088/1367-2630/aabe12} {\bibfield  {journal} {\bibinfo
  {journal} {New Journal of Physics}\ }\textbf {\bibinfo {volume} {20}},\
  \bibinfo {pages} {053010} (\bibinfo {year} {2018})}\BibitemShut {NoStop}%
\bibitem [{\citenamefont {Costa}\ \emph {et~al.}(2018)\citenamefont {Costa},
  \citenamefont {Ringbauer}, \citenamefont {Goggin}, \citenamefont {White},\
  and\ \citenamefont {Fedrizzi}}]{PhysRevA.98.012328}%
  \BibitemOpen
  \bibfield  {author} {\bibinfo {author} {\bibfnamefont {F.}~\bibnamefont
  {Costa}}, \bibinfo {author} {\bibfnamefont {M.}~\bibnamefont {Ringbauer}},
  \bibinfo {author} {\bibfnamefont {M.~E.}\ \bibnamefont {Goggin}}, \bibinfo
  {author} {\bibfnamefont {A.~G.}\ \bibnamefont {White}}, \ and\ \bibinfo
  {author} {\bibfnamefont {A.}~\bibnamefont {Fedrizzi}},\ }\href {\doibase
  10.1103/PhysRevA.98.012328} {\bibfield  {journal} {\bibinfo  {journal} {Phys.
  Rev. A}\ }\textbf {\bibinfo {volume} {98}},\ \bibinfo {pages} {012328}
  (\bibinfo {year} {2018})}\BibitemShut {NoStop}%
\bibitem [{\citenamefont {Helsen}\ \emph {et~al.}(2021)\citenamefont {Helsen},
  \citenamefont {Ioannou}, \citenamefont {Roth}, \citenamefont {Kitzinger},
  \citenamefont {Onorati}, \citenamefont {Werner},\ and\ \citenamefont
  {Eisert}}]{helsen2021estimating}%
  \BibitemOpen
  \bibfield  {author} {\bibinfo {author} {\bibfnamefont {J.}~\bibnamefont
  {Helsen}}, \bibinfo {author} {\bibfnamefont {M.}~\bibnamefont {Ioannou}},
  \bibinfo {author} {\bibfnamefont {I.}~\bibnamefont {Roth}}, \bibinfo {author}
  {\bibfnamefont {J.}~\bibnamefont {Kitzinger}}, \bibinfo {author}
  {\bibfnamefont {E.}~\bibnamefont {Onorati}}, \bibinfo {author} {\bibfnamefont
  {A.~H.}\ \bibnamefont {Werner}}, \ and\ \bibinfo {author} {\bibfnamefont
  {J.}~\bibnamefont {Eisert}},\ }\href {https://arxiv.org/abs/2110.13178}
  {\bibfield  {journal} {\bibinfo  {journal} {arXiv preprint arXiv:2110.13178}\
  } (\bibinfo {year} {2021})}\BibitemShut {NoStop}%
\bibitem [{\citenamefont {Hadfield}\ \emph {et~al.}(2022)\citenamefont
  {Hadfield}, \citenamefont {Bravyi}, \citenamefont {Raymond},\ and\
  \citenamefont {Mezzacapo}}]{hadfield2022measurements}%
  \BibitemOpen
  \bibfield  {author} {\bibinfo {author} {\bibfnamefont {C.}~\bibnamefont
  {Hadfield}}, \bibinfo {author} {\bibfnamefont {S.}~\bibnamefont {Bravyi}},
  \bibinfo {author} {\bibfnamefont {R.}~\bibnamefont {Raymond}}, \ and\
  \bibinfo {author} {\bibfnamefont {A.}~\bibnamefont {Mezzacapo}},\ }\href
  {https://doi.org/10.1007/s00220-022-04343-8} {\bibfield  {journal} {\bibinfo
  {journal} {Communications in Mathematical Physics}\ }\textbf {\bibinfo
  {volume} {391}},\ \bibinfo {pages} {951} (\bibinfo {year}
  {2022})}\BibitemShut {NoStop}%
\bibitem [{\citenamefont {Huang}\ \emph
  {et~al.}(2022{\natexlab{a}})\citenamefont {Huang}, \citenamefont {Kueng},
  \citenamefont {Torlai}, \citenamefont {Albert},\ and\ \citenamefont
  {Preskill}}]{huang2022provably}%
  \BibitemOpen
  \bibfield  {author} {\bibinfo {author} {\bibfnamefont {H.-Y.}\ \bibnamefont
  {Huang}}, \bibinfo {author} {\bibfnamefont {R.}~\bibnamefont {Kueng}},
  \bibinfo {author} {\bibfnamefont {G.}~\bibnamefont {Torlai}}, \bibinfo
  {author} {\bibfnamefont {V.~V.}\ \bibnamefont {Albert}}, \ and\ \bibinfo
  {author} {\bibfnamefont {J.}~\bibnamefont {Preskill}},\ }\href {\doibase
  10.1126/science.abk3333} {\bibfield  {journal} {\bibinfo  {journal}
  {Science}\ }\textbf {\bibinfo {volume} {377}},\ \bibinfo {pages} {eabk3333}
  (\bibinfo {year} {2022}{\natexlab{a}})},\ \Eprint
  {http://arxiv.org/abs/https://www.science.org/doi/pdf/10.1126/science.abk3333}
  {https://www.science.org/doi/pdf/10.1126/science.abk3333} \BibitemShut
  {NoStop}%
\bibitem [{\citenamefont {Elben}\ \emph {et~al.}(2020)\citenamefont {Elben},
  \citenamefont {Kueng}, \citenamefont {Huang}, \citenamefont {van Bijnen},
  \citenamefont {Kokail}, \citenamefont {Dalmonte}, \citenamefont {Calabrese},
  \citenamefont {Kraus}, \citenamefont {Preskill}, \citenamefont {Zoller},\
  and\ \citenamefont {Vermersch}}]{PhysRevLett.125.200501}%
  \BibitemOpen
  \bibfield  {author} {\bibinfo {author} {\bibfnamefont {A.}~\bibnamefont
  {Elben}}, \bibinfo {author} {\bibfnamefont {R.}~\bibnamefont {Kueng}},
  \bibinfo {author} {\bibfnamefont {H.-Y.~R.}\ \bibnamefont {Huang}}, \bibinfo
  {author} {\bibfnamefont {R.}~\bibnamefont {van Bijnen}}, \bibinfo {author}
  {\bibfnamefont {C.}~\bibnamefont {Kokail}}, \bibinfo {author} {\bibfnamefont
  {M.}~\bibnamefont {Dalmonte}}, \bibinfo {author} {\bibfnamefont
  {P.}~\bibnamefont {Calabrese}}, \bibinfo {author} {\bibfnamefont
  {B.}~\bibnamefont {Kraus}}, \bibinfo {author} {\bibfnamefont
  {J.}~\bibnamefont {Preskill}}, \bibinfo {author} {\bibfnamefont
  {P.}~\bibnamefont {Zoller}}, \ and\ \bibinfo {author} {\bibfnamefont
  {B.}~\bibnamefont {Vermersch}},\ }\href {\doibase
  10.1103/PhysRevLett.125.200501} {\bibfield  {journal} {\bibinfo  {journal}
  {Phys. Rev. Lett.}\ }\textbf {\bibinfo {volume} {125}},\ \bibinfo {pages}
  {200501} (\bibinfo {year} {2020})}\BibitemShut {NoStop}%
\bibitem [{\citenamefont {Sarovar}\ \emph {et~al.}(2020)\citenamefont
  {Sarovar}, \citenamefont {Proctor}, \citenamefont {Rudinger}, \citenamefont
  {Young}, \citenamefont {Nielsen},\ and\ \citenamefont
  {Blume-Kohout}}]{Sarovar2020detectingcrosstalk}%
  \BibitemOpen
  \bibfield  {author} {\bibinfo {author} {\bibfnamefont {M.}~\bibnamefont
  {Sarovar}}, \bibinfo {author} {\bibfnamefont {T.}~\bibnamefont {Proctor}},
  \bibinfo {author} {\bibfnamefont {K.}~\bibnamefont {Rudinger}}, \bibinfo
  {author} {\bibfnamefont {K.}~\bibnamefont {Young}}, \bibinfo {author}
  {\bibfnamefont {E.}~\bibnamefont {Nielsen}}, \ and\ \bibinfo {author}
  {\bibfnamefont {R.}~\bibnamefont {Blume-Kohout}},\ }\href {\doibase
  10.22331/q-2020-09-11-321} {\bibfield  {journal} {\bibinfo  {journal}
  {{Quantum}}\ }\textbf {\bibinfo {volume} {4}},\ \bibinfo {pages} {321}
  (\bibinfo {year} {2020})}\BibitemShut {NoStop}%
\bibitem [{\citenamefont {Rudinger}\ \emph {et~al.}(2021)\citenamefont
  {Rudinger}, \citenamefont {Hogle}, \citenamefont {Naik}, \citenamefont
  {Hashim}, \citenamefont {Lobser}, \citenamefont {Santiago}, \citenamefont
  {Grace}, \citenamefont {Nielsen}, \citenamefont {Proctor}, \citenamefont
  {Seritan}, \citenamefont {Clark}, \citenamefont {Blume-Kohout}, \citenamefont
  {Siddiqi},\ and\ \citenamefont {Young}}]{rudinger2021experimental}%
  \BibitemOpen
  \bibfield  {author} {\bibinfo {author} {\bibfnamefont {K.}~\bibnamefont
  {Rudinger}}, \bibinfo {author} {\bibfnamefont {C.~W.}\ \bibnamefont {Hogle}},
  \bibinfo {author} {\bibfnamefont {R.~K.}\ \bibnamefont {Naik}}, \bibinfo
  {author} {\bibfnamefont {A.}~\bibnamefont {Hashim}}, \bibinfo {author}
  {\bibfnamefont {D.}~\bibnamefont {Lobser}}, \bibinfo {author} {\bibfnamefont
  {D.~I.}\ \bibnamefont {Santiago}}, \bibinfo {author} {\bibfnamefont {M.~D.}\
  \bibnamefont {Grace}}, \bibinfo {author} {\bibfnamefont {E.}~\bibnamefont
  {Nielsen}}, \bibinfo {author} {\bibfnamefont {T.}~\bibnamefont {Proctor}},
  \bibinfo {author} {\bibfnamefont {S.}~\bibnamefont {Seritan}}, \bibinfo
  {author} {\bibfnamefont {S.~M.}\ \bibnamefont {Clark}}, \bibinfo {author}
  {\bibfnamefont {R.}~\bibnamefont {Blume-Kohout}}, \bibinfo {author}
  {\bibfnamefont {I.}~\bibnamefont {Siddiqi}}, \ and\ \bibinfo {author}
  {\bibfnamefont {K.~C.}\ \bibnamefont {Young}},\ }\href {\doibase
  10.1103/PRXQuantum.2.040338} {\bibfield  {journal} {\bibinfo  {journal} {PRX
  Quantum}\ }\textbf {\bibinfo {volume} {2}},\ \bibinfo {pages} {040338}
  (\bibinfo {year} {2021})}\BibitemShut {NoStop}%
\bibitem [{\citenamefont {Milz}\ and\ \citenamefont
  {Modi}(2021)}]{Milz2021PRXQ}%
  \BibitemOpen
  \bibfield  {author} {\bibinfo {author} {\bibfnamefont {S.}~\bibnamefont
  {Milz}}\ and\ \bibinfo {author} {\bibfnamefont {K.}~\bibnamefont {Modi}},\
  }\href {\doibase 10.1103/PRXQuantum.2.030201} {\bibfield  {journal} {\bibinfo
   {journal} {PRX Quantum}\ }\textbf {\bibinfo {volume} {2}},\ \bibinfo {pages}
  {030201} (\bibinfo {year} {2021})},\ \Eprint
  {http://arxiv.org/abs/2012.01894} {arXiv:2012.01894} \BibitemShut {NoStop}%
\bibitem [{\citenamefont {Huang}\ \emph
  {et~al.}(2022{\natexlab{b}})\citenamefont {Huang}, \citenamefont {Broughton},
  \citenamefont {Cotler}, \citenamefont {Chen}, \citenamefont {Li},
  \citenamefont {Mohseni}, \citenamefont {Neven}, \citenamefont {Babbush},
  \citenamefont {Kueng}, \citenamefont {Preskill},\ and\ \citenamefont
  {McClean}}]{doi:10.1126/science.abn7293}%
  \BibitemOpen
  \bibfield  {author} {\bibinfo {author} {\bibfnamefont {H.-Y.}\ \bibnamefont
  {Huang}}, \bibinfo {author} {\bibfnamefont {M.}~\bibnamefont {Broughton}},
  \bibinfo {author} {\bibfnamefont {J.}~\bibnamefont {Cotler}}, \bibinfo
  {author} {\bibfnamefont {S.}~\bibnamefont {Chen}}, \bibinfo {author}
  {\bibfnamefont {J.}~\bibnamefont {Li}}, \bibinfo {author} {\bibfnamefont
  {M.}~\bibnamefont {Mohseni}}, \bibinfo {author} {\bibfnamefont
  {H.}~\bibnamefont {Neven}}, \bibinfo {author} {\bibfnamefont
  {R.}~\bibnamefont {Babbush}}, \bibinfo {author} {\bibfnamefont
  {R.}~\bibnamefont {Kueng}}, \bibinfo {author} {\bibfnamefont
  {J.}~\bibnamefont {Preskill}}, \ and\ \bibinfo {author} {\bibfnamefont
  {J.~R.}\ \bibnamefont {McClean}},\ }\href {\doibase 10.1126/science.abn7293}
  {\bibfield  {journal} {\bibinfo  {journal} {Science}\ }\textbf {\bibinfo
  {volume} {376}},\ \bibinfo {pages} {1182} (\bibinfo {year}
  {2022}{\natexlab{b}})},\ \Eprint
  {http://arxiv.org/abs/https://www.science.org/doi/pdf/10.1126/science.abn7293}
  {https://www.science.org/doi/pdf/10.1126/science.abn7293} \BibitemShut
  {NoStop}%
\bibitem [{\citenamefont {Huang}\ \emph {et~al.}(2021)\citenamefont {Huang},
  \citenamefont {Kueng},\ and\ \citenamefont
  {Preskill}}]{PhysRevLett.126.190505}%
  \BibitemOpen
  \bibfield  {author} {\bibinfo {author} {\bibfnamefont {H.-Y.}\ \bibnamefont
  {Huang}}, \bibinfo {author} {\bibfnamefont {R.}~\bibnamefont {Kueng}}, \ and\
  \bibinfo {author} {\bibfnamefont {J.}~\bibnamefont {Preskill}},\ }\href
  {\doibase 10.1103/PhysRevLett.126.190505} {\bibfield  {journal} {\bibinfo
  {journal} {Phys. Rev. Lett.}\ }\textbf {\bibinfo {volume} {126}},\ \bibinfo
  {pages} {190505} (\bibinfo {year} {2021})}\BibitemShut {NoStop}%
\bibitem [{\citenamefont {Aloisio}\ \emph {et~al.}(2022)\citenamefont
  {Aloisio}, \citenamefont {White}, \citenamefont {Hill},\ and\ \citenamefont
  {Modi}}]{aloisio-complexity}%
  \BibitemOpen
  \bibfield  {author} {\bibinfo {author} {\bibfnamefont {I.~A.}\ \bibnamefont
  {Aloisio}}, \bibinfo {author} {\bibfnamefont {G.~A.~L.}\ \bibnamefont
  {White}}, \bibinfo {author} {\bibfnamefont {C.~D.}\ \bibnamefont {Hill}}, \
  and\ \bibinfo {author} {\bibfnamefont {K.}~\bibnamefont {Modi}},\ }\href@noop
  {} {\bibfield  {journal} {\bibinfo  {journal} {In preparation}\ } (\bibinfo
  {year} {2022})}\BibitemShut {NoStop}%
\bibitem [{\citenamefont {Luchnikov}\ \emph {et~al.}(2020)\citenamefont
  {Luchnikov}, \citenamefont {Vintskevich}, \citenamefont {Grigoriev},\ and\
  \citenamefont {Filippov}}]{PhysRevLett.124.140502}%
  \BibitemOpen
  \bibfield  {author} {\bibinfo {author} {\bibfnamefont {I.~A.}\ \bibnamefont
  {Luchnikov}}, \bibinfo {author} {\bibfnamefont {S.~V.}\ \bibnamefont
  {Vintskevich}}, \bibinfo {author} {\bibfnamefont {D.~A.}\ \bibnamefont
  {Grigoriev}}, \ and\ \bibinfo {author} {\bibfnamefont {S.~N.}\ \bibnamefont
  {Filippov}},\ }\href {\doibase 10.1103/PhysRevLett.124.140502} {\bibfield
  {journal} {\bibinfo  {journal} {Phys. Rev. Lett.}\ }\textbf {\bibinfo
  {volume} {124}},\ \bibinfo {pages} {140502} (\bibinfo {year}
  {2020})}\BibitemShut {NoStop}%
\bibitem [{\citenamefont {Cerrillo}\ and\ \citenamefont
  {Cao}(2014)}]{PhysRevLett.112.110401}%
  \BibitemOpen
  \bibfield  {author} {\bibinfo {author} {\bibfnamefont {J.}~\bibnamefont
  {Cerrillo}}\ and\ \bibinfo {author} {\bibfnamefont {J.}~\bibnamefont {Cao}},\
  }\href {\doibase 10.1103/PhysRevLett.112.110401} {\bibfield  {journal}
  {\bibinfo  {journal} {Phys. Rev. Lett.}\ }\textbf {\bibinfo {volume} {112}},\
  \bibinfo {pages} {110401} (\bibinfo {year} {2014})}\BibitemShut {NoStop}%
\bibitem [{\citenamefont {Figueroa-Romero}\ \emph {et~al.}(2021)\citenamefont
  {Figueroa-Romero}, \citenamefont {Modi}, \citenamefont {Harris},
  \citenamefont {Stace},\ and\ \citenamefont {Hsieh}}]{PRXQuantum.2.040351}%
  \BibitemOpen
  \bibfield  {author} {\bibinfo {author} {\bibfnamefont {P.}~\bibnamefont
  {Figueroa-Romero}}, \bibinfo {author} {\bibfnamefont {K.}~\bibnamefont
  {Modi}}, \bibinfo {author} {\bibfnamefont {R.~J.}\ \bibnamefont {Harris}},
  \bibinfo {author} {\bibfnamefont {T.~M.}\ \bibnamefont {Stace}}, \ and\
  \bibinfo {author} {\bibfnamefont {M.-H.}\ \bibnamefont {Hsieh}},\ }\href
  {\doibase 10.1103/PRXQuantum.2.040351} {\bibfield  {journal} {\bibinfo
  {journal} {PRX Quantum}\ }\textbf {\bibinfo {volume} {2}},\ \bibinfo {pages}
  {040351} (\bibinfo {year} {2021})}\BibitemShut {NoStop}%
\bibitem [{\citenamefont {Wood}\ \emph {et~al.}(2011)\citenamefont {Wood},
  \citenamefont {Biamonte},\ and\ \citenamefont {Cory}}]{wood2011tensor}%
  \BibitemOpen
  \bibfield  {author} {\bibinfo {author} {\bibfnamefont {C.~J.}\ \bibnamefont
  {Wood}}, \bibinfo {author} {\bibfnamefont {J.~D.}\ \bibnamefont {Biamonte}},
  \ and\ \bibinfo {author} {\bibfnamefont {D.~G.}\ \bibnamefont {Cory}},\
  }\href {https://arxiv.org/abs/1111.6950} {\bibfield  {journal} {\bibinfo
  {journal} {arXiv preprint arXiv:1111.6950}\ } (\bibinfo {year}
  {2011})}\BibitemShut {NoStop}%
\end{thebibliography}

\clearpage
\appendix
\onecolumngrid

\section{Process tensors and process tensor marginals from classical shadow tomography}
\label{app:process-tensors}
We review here the background of process tensors, and detail the procedure to obtain process tensor marginals from the classic shadow data. For a more extensive discussion, see Refs.~\cite{Pollock2018a,Milz2021PRXQ,white2022non}.

\emph{Process tensor framework.--}
A quantum stochastic process describes the non-deterministic dynamics of an open quantum system $\rho^S\in \mathcal{B}(\mathcal{H}_S)$ across a series of times $\mathbf{T}_k=\{t_0,t_1,\cdots,t_k\}$. Specifically, this represents knowledge of the conditional state $\rho_k^S(\mathbf{A}_{k-1:0})$ subject to an arbitrary sequence $\mathbf{A}_{k-1:0}:= \{\mathcal{A}_0, \mathcal{A}_1, \cdots, \mathcal{A}_{k-1}\}$ of control operations at each $t_j$. Note that these controls in full generality constitute non-deterministic maps, such as quantum instruments.
%we consider the situation where a $k$-step process is driven by a sequence $\mathbf{A}_{k-1:0}$ of control operations, each represented mathematically by CP maps: $\mathbf{A}_{k-1:0} := \{\mathcal{A}_0, \mathcal{A}_1, \cdots, \mathcal{A}_{k-1}\}$, after which we obtain a final state $\rho_k(\mathbf{A}_{k-1:0})$ conditioned on this choice of interventions. 
These controlled dynamics have the form:
\begin{equation}\label{eq:multiproc-app}
        \rho_k^S\left(\textbf{A}_{k-1:0}\right) = \text{tr}_E [U_{k:k-1} \, \mathcal{A}_{k-1} \cdots \, U_{1:0} \, \mathcal{A}_{0} (\rho^{SE}_0)],
\end{equation}
where $U_{k:k-1}(\cdot) = u_{k:k-1} (\cdot) u_{k:k-1}^\dag$. Eq.~\eqref{eq:multiproc-app} can be used to define a mapping from past controls $\mathbf{A}_{k-1:0}$ to future states $\rho_k\left(\textbf{A}_{k-1:0}\right)$, which is the process tensor $\mathcal{T}_{k:0}$:
\begin{equation}
\label{eq:PT}
    \mathcal{T}_{k:0}\left[\mathbf{A}_{k-1:0}\right] = \rho_k^S(\mathbf{A}_{k-1:0}).
\end{equation}
Each intervention $\mathcal{A}_j$ has a Choi representation $\hat{\mathcal{A}}_j$ via the Choi-Jamiolkowski isomorphism (CJI), given by its action on one half of a maximally entangled state $\ket{\Phi^+} = \sum_{i=1}^d\ket{ii}$:
\begin{equation}
\label{eq:channel-choi}
    \hat{\mathcal{A}}_j = (\mathcal{A}\otimes \mathcal{I})\left[|\Phi^+\rangle\!\langle \Phi^+|\right] = \sum_{i,j=1}^d\mathcal{A}[|i\rangle\!\langle j|]\otimes |i\rangle\!\langle j|.
\end{equation}
The two subsystems of this state can be seen as `input' and `output' spaces: $\hat{\mathcal{A}}\in \mathcal{B}(\mathcal{H}_{\mathfrak{o}})\otimes \mathcal{B}(\mathcal{H}_{\mathfrak{i}})$, since the output state under the action of the map is given on the left space after projecting some state onto the right subsystem. That is, $\mathcal{A}[\sigma] = \text{Tr}_{\mathfrak{i}}\left[(\mathbb{I}_{\mathfrak{o}}\otimes \sigma^{\text{T}})\hat{\mathcal{A}}\right]$.

Through a generalisation of the CJI, process tensors may be represented as a $2k+1$-partite quantum state, $\Upsilon_{k:0}$. Equation~\eqref{eq:channel-choi} is generalised by swapping in one half of a fresh maximally entangled state at each time. This state then predicts multi-time probabilities in accordance with Equation~\eqref{eq:multiproc-app}. That is, when the final state $\rho_k^S(\mathbf{A}_{k-1:0})$ is measured with a POVM $\{\Pi_k\}$, the observed probabilities for each effect is given by
\begin{equation}
\label{eq:PT-vector}
    p_k^S(\mathbf{A}_{k-1:0})= \text{Tr}  \left[\Upsilon_{k:0}\left(\Pi_k\otimes \hat{\mathcal{A}}_{k-1}\otimes \cdots \otimes  \hat{\mathcal{A}}_0\right)^\text{T}\right].
\end{equation}
Thus, sequences of CP maps constitute observables on the process tensor. If a process output is measured when conditioned on an informationally complete set of instruments, then Equation~\eqref{eq:PT-vector} can be uniquely inverted to reconstruct $\Upsilon_{k:0}$ -- although in practice when the probabilities are actually noisy frequency estimates, an estimation procedure such as maximum likelihood must be employed to ensure physical conditions are imposed, like positivity of the map and causal direction of the statistics.

% \begin{equation}
% \label{PT-vector-est}
%     p_{i,\vec{\mu}} = \text{Tr}\left[(\Pi_i \otimes  \mathcal{B}_k^{\mu_{k-1}\text{T}}\otimes \cdots \otimes \mathcal{B}_0^{\mu_0\text{T}})\Upsilon_{k:0}\right].
% \end{equation}

The Choi form of a process tensor is an $2k+1$-partite state with process marginals, $\{\hat{\mathcal{E}}_{k:k-1},\cdots,\hat{\mathcal{E}}_{2:1},\hat{\mathcal{E}}_{1:0},\rho_0\}$. Any correlations between these subsystems constitute non-Markovian correlations due to the external environment. In keeping with our channel Choi state convention, we use the notation $\mathfrak{o}_j$ to denote an output leg of the process at time $t_j$, and $\mathfrak{i}_j$ for the input leg of the process at time $t_{j-1}$. The collection of indices is therefore $\{\mathfrak{o}_k,\mathfrak{i}_k,\cdots,\mathfrak{o}_2,\mathfrak{i}_2,\mathfrak{o}_1,\mathfrak{i}_1,\mathfrak{o}_0\}$. 

Assuming now that we have a process tensor defined on a register $\mathbf{Q}=\{q_1,\cdots,q_N\}$, then we can fine-grain not just in time but also in space. We first note that without loss of informational completeness, each instrument $\hat{\mathcal{A}}_j$ can be factored into local operations $\bigotimes_{i=1}^N\hat{\mathcal{A}}_j^{(q_i)}$. The dynamical maps of the process can also be further marginalised in the same way to obtain $\hat{\mathcal{E}}_{j:j-1}^{(q_i)}$: the average CPTP map taking the $i$th qubit from time $t_{j-1}$ to $t_j$. Although process tensors contain information about all multi-time correlations, for simplicity's (and efficiency's) sake, in this work we consider only two map correlations. Therefore, spacetime correlations can be probed across the following pairs of marginals:

\begin{itemize}
    \item \underline{Purely spatial}: $\hat{\mathcal{E}}_{j:j-1}^{(q_n)}$ and $\hat{\mathcal{E}}_{j:j-1}^{(q_m)}$,
    \item \underline{Purely temporal}: $\hat{\mathcal{E}}_{j:j-1}^{(q_n)}$ and $\hat{\mathcal{E}}_{l:l-1}^{(q_n)}$,
    \item \underline{Spatiotemporal}: $\hat{\mathcal{E}}_{j:j-1}^{(q_n)}$ and $\hat{\mathcal{E}}_{l:l-1}^{(q_m)}$.
\end{itemize}

Pure spatial correlations constitute either direct crosstalk, or presence of an environmental common cause factor. As we show in the next section, purely temporal marginals constitute genuine bath non-Markovian correlations. And finally, spatiotemporal correlations indicate two qubits coupled to the same non-Markovian bath.

\emph{Recovering process marginals.--}

Although collecting shadows suffices to estimate properties of the process marginals, they will not in general be able to reconstruct a physical estimate of the process marginals. To this effect we estimate enough observables to constitute informationally complete information about the process tensor, and then employ maximum-likelihood PTT to process the data and obtain a physical estimate. The advantages of a physical (positive, causal) estimate is that we make use of information theoretic tools that rely on the postitivity of the state. 

The total number of shadows required to estimate $M$ observables $O_i$ to a precision of $\epsilon$ was shown by Huang et al. to be~\cite{huang-shadow}
\begin{equation}
    N_{\text{shots}} = \mathcal{O}\left(\frac{\log M}{\epsilon^2} \max_{1\leq i \leq M}\| O_i\|_{\text{shadow}}\right).
\end{equation}
For an $l$-local Pauli observable, this shadow norm is shown to be $3^l$. The numerical simulations in the main text are two step processes, whose Choi forms are equivalent to a five qubit state. The remainder of the observables on the process are trivial, i.e. $\mathbb{I}$. In the Choi picture, this is equivalent to a maximally depolarising channel. Explicitly,
\begin{equation}
    \mathcal{R}_\lambda[\rho] := (1 - \lambda)\rho + \lambda\mathbb{I}/2,
\end{equation}
then $\hat{\mathcal{R}}_{1} = \mathbb{I}/2$, with normalisation chosen to be $d$.

Causality conditions fix the expectations of $\sum_{j=1}^k(d^2-1)d^{4j-2}$ Pauli operators. For a two step process then, this leaves $M=820$ expectation values to estimate. From the bounds given in Ref.~\cite{huang-shadow}, we therefore have $N_{\text{shots}} = \mathcal{O}(\log(N_{\text{qubits}}\cdot M)/\epsilon^2 3^5)\approx 2\times 10^7$ measurements in the worst case, which is what we use in our numerical experiments to achieve a precision of $\epsilon=0.01$. The median-of-means algorithm from Ref.~\cite{huang-shadow} is used to estimate an informationally complete set of observables for each process marginal, and the maximum likelihood estimation obtained using the methods developed in Ref.~\cite{white2022non}.

% An estimate for the map is coupled with metric of goodness (the likelihood) which quantifies how consistent the map is with the data. The cost function is then minimised while enforcing the physicality of the map.
% The stored data vector in PTT is the object $n_{i,\vec{\mu}}$, which contains the observed measurement probabilities for the $i$th effect of an IC-POVM, subject to a sequence of $k$ operations $\bigotimes_{j=0}^{k-1}\mathcal{B}_j^{\mu_j}$. As is typical in MLE tomography, this data is fit to a model for the process, $\Upsilon_{k:0}$, such that
% \begin{equation}
% \label{PT-vector-est}
%     p_{i,\vec{\mu}} = \text{Tr}\left[(\Pi_i \otimes  \mathcal{B}_k^{\mu_{k-1}\text{T}}\otimes \cdots \otimes \mathcal{B}_0^{\mu_0\text{T}})\Upsilon_{k:0}\right].
% \end{equation}
% These predictions are then compared to the observed frequencies, $n_{i,\vec{\mu}}$. The `likelihood' of $\Upsilon_{k:0}$ subject to the data is given by $\mathcal{L} = \prod_{i,\vec{\mu}} (p_{i,\vec{\mu}})^{n_{i,\vec{\mu}}}$. The cost function of MLE algorithms is then the log-likelihood, i.e.,
% \begin{equation}
% \label{ML-cost}
%     f(\Upsilon_{k:0}) = -\ln \mathcal{L} = \sum_{i,\vec{\mu}} -n_{i,\vec{\mu}}\ln p_{i,\vec{\mu}},
% \end{equation}

\section{Pathways proof}
\label{app:pathways-proof}
\emph{Distinguishing RNM and BNM.--}
Here, we give a short graphical proof for the fact that applying depolarising channels to other qubits in the register at each time eliminates them as a potential non-Markovian source.
\begin{figure}
    \centering
    \includegraphics[width=\linewidth]{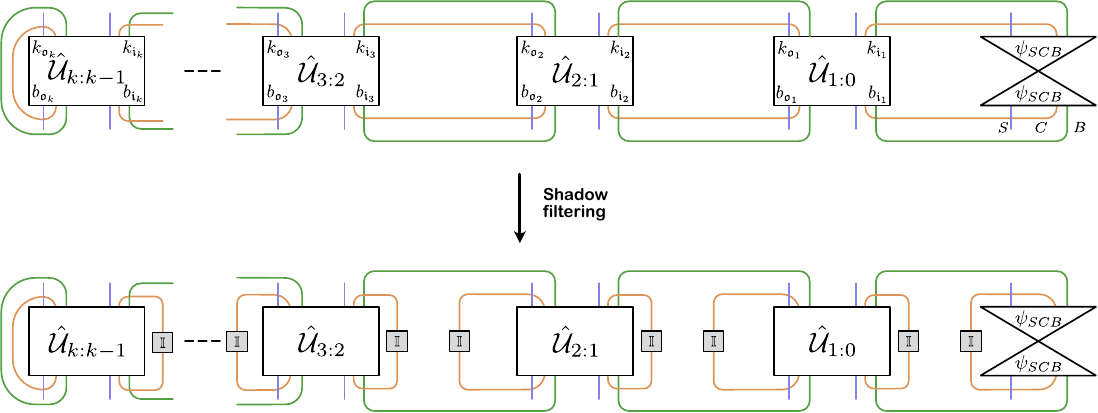}
    \caption{Graphical depiction of causal testing. At the top, we represent the process tensor in terms of its dilated link product representation. Below, we show how the transformation implements depolarising channels, leaving the process as tensor product structure across $\mathcal{B}(\mathcal{H}_C)$.}
    \label{fig:PT-link}
\end{figure}
In the dilated open quantum evolution picture, a process tensor may be written as the \emph{link product} $\star$ of a series of $SE$ unitaries. 
\begin{equation}
\label{eq:PT-link}
\Upsilon_{k:0} = \text{Tr}_{E}\left[\bigstar_{i=1}^{k} \hat{\mathcal{U}}^{(SE)}_{i:i-1} \star \rho_0^{(SE)}\right],
\end{equation}
where $\hat{\mathcal{U}}_{i:i-1}$ is the Choi representation of the $i$th step unitary $U_{i:i-1}$. This lives on $\mathcal{B}(\mathcal{H})_S^{\mathfrak{i}}\otimes\mathcal{B}(\mathcal{H})_E^{\mathfrak{i}}\otimes\mathcal{B}(\mathcal{H})_S^{\mathfrak{o}}\otimes\mathcal{B}(\mathcal{H})_E^{\mathfrak{o}}$. Suppose the environment factorises into two spaces: $\mathcal{H}_E \cong \mathcal{H}_C \otimes \mathcal{H}_B$ -- the controllable qubits and the uncontrollable bath -- then each $\mathcal{U}_{i:i-1}$ has twelve indices: ket and bra for the input and output spaces for each of $S, C, B$. Equation~\eqref{eq:PT-link} can be written more intuitively using the graphical calculus formalism~\cite{wood2011tensor}, as shown at the top of Figure~\ref{fig:PT-link}.

% \begin{equation}
% \Upsilon_{k:0} = \sum \delta_{k_k^ck_k^b}^{b_k^cb_k^b} \left(\mathcal{U}_{k:k-1}\right)^{k^s_kk^c_kk^b_k}_{b^s_kb^c_kb^b_k} \left(\mathcal{U}_{k-1:k-2}\right)^{b^s_kb^c_kb^b_k}_{k^s_{k-1}k^c_{k-1}k^b_{k-1}}\cdots \left(\mathcal{U}_{1:k0}\right)^{b^s_kb^c_kb^b_k}_{k^s_{k-1}k^c_{k-1}k^b_{k-1}}
% \end{equation}

Because depolarising channels are a product state over $\mathcal{B}(\mathcal{H}_{\mathfrak{o}})\otimes \mathcal{B}(\mathcal{H}_{\mathfrak{i}})$, they hence implement a causal break.
Note that any choice of instrument for which the Choi state is a product state (for example, a projective measurement and fresh preparation) constitutes a valid causal break, but a depolarising channel is the only one for which the locality of the observable does not grow, and hence can be efficient under the classical shadows procedure.

\emph{Identifying systems with a shared non-Markovian bath.--}
The same logic applies for the observation of correlations between some $\hat{\mathcal{E}}_{j:j-1}^{(q_n)}$ and $\hat{\mathcal{E}}_{l:l-1}^{(q_m)}$. Obtaining these two correlated marginals means that a causal break is applied at the $\mathfrak{o}$ leg of $\hat{\mathcal{E}}_{j:j-1}^{(q_m)}$ and at the $\mathfrak{i}$ leg of $\hat{\mathcal{E}}_{l:l-1}^{(q_n)}$. Hence, this erases any mixing terms between $\mathcal{H}_{q_n}$ and $\mathcal{H}_{q_m}$ in the unitaries. Consequently, non-zero values for the mutual information between these maps must be due to interactions involving $\mathcal{H}_B$ \& $\mathcal{H}_{q_n}$ at the first time step, and then $\mathcal{H}_B$ \& $\mathcal{H}_{q_m}$ at the second second. I.e. the two qubits shared the same bath. 

%\com{There should be a picture to go here for both cases. An a/b link product, one for temporal correlations and one for spatiotemporal ones}
%\com{Explicitly state that this gives a sense of why temporal shadows are quite restrictive in this scenario}

% \section{Entropic inequalities applied to source testing}
% \label{app:entropic-inequalities}

% \begin{itemize}
%     \item We have a 2D quantum state from which we are computing marginals. 
% \end{itemize}

% \section{Illustrative examples and non-examples}
% \label{app:examples}
% Do some relatively simple circuit diagrams (maybe where we can analytically write down the PT). This can include looking at temporal shadows with (non-)unital dynamics, as well as looking at the shared bath case where a register qubit is a common cause influence on the bath and the system.

\end{document}